# Meeting the Embedded Design Needs of Automotive Applications

Wayne Lyons, ARM

**Abstract**

*The importance of embedded systems in driving innovation in automotive applications continues to grow. Understanding the specific needs of developers targeting this market is also helping to drive innovation in RISC core design. This paper describes how a RISC instruction set architecture has evolved to better meet those needs, and the key implementation features in two very different RISC cores are used to demonstrate the challenges of designing for real-time automotive systems.*

## 1 Introduction

The dominant trend in RISC core design has been to follow a path towards higher CPU clock frequencies and more complex feature sets. This 'up-and-to-the-right' industry roadmap is certain to continue, with many embedded applications requiring support for ever higher levels of performance and richer functionality. Automotive embedded functions, however, are characterized by a mixture of high-performance systems as well as far simpler, low-performance but extremely cost-sensitive applications. Clearly, it is not possible to satisfy these opposite ends of the performance spectrum with the same physical device.

Currently, a typical automotive system consists of a physically distributed network of 8/16-bit and 32-bit processors that are dedicated to specific tasks within the car. These tasks range from relatively simple and straightforward functions such as monitoring an oil pressure sensor or controlling the movement of electric windows, to much more complex and demanding engine management, in car entertainment or navigation systems. The simpler tasks are typically addressed by the use of 8 and 16-bit microcontrollers.

The automotive industry's vision for the future is that these discrete and distributed compute resources could be collectively harnessed to enable application processing requirements to be distributed on what is effectively a powerful virtual multi-core processing system. To enable this vision, one of the key requirements will be the harmonization of the instruction set architecture (ISA) throughout the entire automotive system. Apart from the ability to distribute tasks, designing to a common ISA will also deliver benefits to the manufacturers in terms of the ability to reuse code, and broader development economies of scale.

A precondition for any automotive manufacturer to be able standardize on a particular ISA is the availability of processor implementations that will satisfy efficiently both ends of the performance spectrum. This demands an architecture that has sufficient flexibility to enable the design of very small and efficient 32-bit processors that rival 8-bit controllers in cost-effectiveness, as well as high-frequency, high-performance processor cores.

## 2 ISA Foundation

The ARM architecture has been extremely successful in addressing a broad spectrum of embedded design requirements, and has evolved over a number of years to provide a foundation for a portfolio of core implementations offered by many different manufacturers. ARM is now the most widely used architecture for new embedded designs.

One of the key features that has driven the popularity of the architecture has been the compressed instruction set extension known as 'Thumb®'. The Thumb instruction set features a subset of the most commonly used 32-bit ARM instructions, which have been compressed into 16-bit wide opcodes. On execution, these 16-bit instructions are decompressed transparently to full 32-bit ARM instructions in real time without performance loss. In analysing the needs of automotive developers, it became clear that historically Thumb has been, and continues to be used extensively to achieve the best possible code density. Achieving good code density implies minimum memory footprint, which is critical in reducing system cost.

It became apparent that an evolution of Thumb – known as Thumb-2 technology – could enhance the performance and code density even further to the benefit of all embedded applications. The new Thumb-2 ISA extension has provided the capability for several other feature improvements introduced in the latest ARM core implementations.

### 2.1 Thumb-2 Technology Development

Memory dominates the bill of materials cost in virtually all embedded automotive applications. As well as improving performance over the original Thumb specification, Thumb-2 technology seeks to maximize usage of cache, as well as tightly coupled and embedded flash memory, and so helps to reduce the overall memory footprint.

Thumb-2 technology is a blend of 16 and 32-bit instructions. In developing Thumb-2 technology, the initial objectives were to achieve the performance of 32-bit ARM instructions yet match the code density of the original Thumb instruction set. Preliminary EEMBC benchmarking data substantiates the original performance objective.

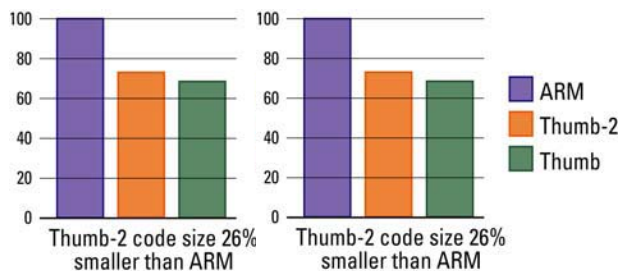

Figure 1: Thumb-2 Performance and Code Size

To improve performance over the original Thumb implementation, Thumb-2 technology has incorporated some of the most popular ARM instructions for data processing, DSP and media functionality, as well as the load and store multiple instructions that are used for context switching in and out of subroutines. Conditional branch is supported to enable longer branching instructions and support a bigger target address range – effectively allowing a branch anywhere within the target memory address range without having to go through an indirection.

Thumb-2 technology also supports the ability to add a co-processor, if required. The Vector Floating Point (VFP) co-processor is becoming relevant to automotive applications since it allows designers to work directly with floating point output from popular code generation tools, enabling a smooth migration to autocode generation. Floating point capability also enables the developer to work in the native units of measurement without the need for scaling operations. Often, this route is followed for convenience and to benefit from a more efficient process in translating the output from automatic code generation tools, rather than conferring any performance benefit within the application.

For developers who wish to scale values read from sensors, the hardware divide instruction enables data to be scaled very efficiently.

Thumb-2 technology also incorporates completely new functionality, specifically aimed at supporting deeply embedded control applications. Deeply embedded applications are characterised by the need to process large volumes of general-purpose input-output (I/O) data. Often, the core requirement is to extract bit-level data from specific pins or sensors, which may have been brought in to the device as an 8- or 16-bit word. In order to utilize that information, the required bits must be extracted. In many legacy 8-bit development environments, bit-manipulation instructions are provided but implemented in micro-code they typically take several cycles to execute a simple bit set or bit clear function.

With Thumb-2 technology, support for bit manipulation through bit field insert and clear, as well as bit reversal instructions, allow comprehensive bit manipulation to be performed very efficiently. In previous RISC architectures, typically the whole word would be read in and then rotated over several cycles to get the required bit into the correct location to perform the logical manipulation. The new Thumb-2 instructions achieve this operation by representing the information to the processor as bits within the whole port, enabling the bit-level information to be extracted and manipulated much more efficiently with direct instructions.

**2.2 Supporting Flash Memory**
Embedded flash is an important and popular technology in many embedded applications, providing flexibility and continuity in data storage. However, embedded flash memory is slow, typically operating at only 30-40MHz. As processor cores and digital logic has benefited from increased performance through architectural and technology enhancements, embedded memory performance has not moved forward significantly. For example, the expected performance of flash memory in 0.25-micron process would typically have been around 40MHz. In moving to a 0.18-micron process this might only increase to around 41MHz. Consequently, there has been a growing gap between the CPU speed and the memory speed.

To accommodate this performance gap, in designing the flash memory interface designers typically fetch more than one 16-bit value at a time – effectively streaming data from the memory to keep the processor busy.

In moving data, if the data cannot be moved immediately with the instruction, a literal pool fetch is typically used. This pool of immediate data is held quite close to where the program counter is currently located. Consequently, literal pools become spread throughout the memory map. This means that the fetch sequence from the flash memory can be disrupted by a non-sequential fetch of data from the literal pool, before returning back to the sequential fetch. The stream of information is therefore broken from the embedded flash, and consequently, performance is degraded. Benchmarks show a performance degradation of 15 percent is possible because of this effect. This non-sequential fetch of data which occurs when the instruction stream breaks the sequential access, imposes significant performance degradation. A cached architecture will typically outperform a Harvard machine by a similar margin under these conditions.

The 16-bit immediate instructions MOVW and MOVH help to address this performance degradation. The literal pool is loaded immediately within the instruction, by effectively sequentially loading the data with the instruction with no break to go off and fetch from the literal pool. This mechanism restores the sequential nature of instruction accesses being made to the flash.

**2.3 Targeting C**

The use of C as a high-level language is almost universal for deeply embedded application environments, so ensuring that the instruction set is as efficient as possible as a C target is very important. Adding two new instructions that compile very efficiently from C, help to improve both performance and memory footprint.

The conditional 'IF-THEN', or IT, instruction carries with it a set of conditional codes that predicate the execution of the instructions that follow. In a Thumb environment, conditional branching is addressed by having multiple branches. In Thumb-2, having a conditional IT instruction means that it is possible to step straight through the conditional code and the instructions are executed (or not executed – represented by a 'NOP') instead of having to branch too indiscriminately. Again, this instruction encourages sequencing of opcodes rather than branching, which helps to maintain processor performance.

The Table branch instruction is useful for implementing switch case statements where the conditional test is made at the beginning and a sequence of code is executed conditionally.

| Processor Core | Scaled GM/MHz | |
|---|---|---|
| ARM7™ (ARM) | 28453.8 | (100%) |
| ARM7 (Thumb) | 22527.8 | (79%) |
| **Cortex-M3™ (Thumb-2)** | **38899.2** | **(137%)** |

| Processor Core | Code Size | |
|---|---|---|
| ARM7 (ARM) | 21168 | (100%) |
| Thumb (Thumb) | 12106 | (57%) |
| **Cortex-M3 (Thumb-2)** | **12106** | **(57%)** |

Table 1: Comparing Thumb-2 performance and code density with Thumb and ARM

Table 1 provides preliminary benchmark data for a Thumb-2 implementation. Note that scaling is the Geometric Mean of the 6 available AutoIndy benchmarks for these cores. Preliminary the ARM Cortex™-M3 processor figures developed on AutoIndy benchmarks. Since these are preliminary benchmark figures, further improvements can be expected for final data.

## 3 Thumb-2 Implementations

The flexibility of Thumb-2 technology in supporting the needs of the automotive market is demonstrated in two distinct core implementations. The ARM1156T2F-S™ processor is a high-performance design intended for high-end applications, which includes, among many other new features, a fault-tolerant memory interface and fine-grained memory protection unit (MPU). The new ARM Cortex- M3 processor is a silicon-efficient 32-bit processor that is cost-competitive with 8- and 16-bit devices.

### 3.1 ARM1156T2F-S Processor

The features incorporated into the design of the ARM1156T2F-S processor core have been specified after extensive consultation with automotive companies and industry bodies. Particular attention has been paid to the requirements of OSEK (Version 2.1.1) compliant real-time operating systems (RTOS). OSEK is a European automotive industry standards effort to produce open systems interfaces for vehicle electronics.

#### 3.1.1 Fine-grain Memory Protection Unit

A key part of OSEK's strategic vision is to enable re-usability of software modules in order to deliver economies of scale and time-to-market benefits to the industry. Automotive component suppliers want to be able to select software from partners to most effectively address the needs of specific functions. For example, in the engine domain, a manufacturer may benefit from being able to re-use a third-party software model to characterise engine wear and tear. In order to enable software re-use from several different sources, it is vital that there is a mechanism to allow each module to be isolated so that the risk of interference is minimised. The requirement is to be able to isolate or 'lock down' many of these routines.

Current Memory Protection Units (MPUs) typically offer 4KByte code boundaries for isolating functions. This is typically too large for systems which have limited memory resource. Large code boundaries mean that it is impossible to efficiently segregate many small software routines – often several tasks will have to be included within the same protection scheme.

To enable good segregation of individual tasks executed within an OS kernel, the new ARM MPU has been completely re-engineered to provide a finer granularity of memory region for each task. This enables more effective use of the available memory resource for supporting multiple tasks. This MPU architecture will be carried forward to future ARM cores.

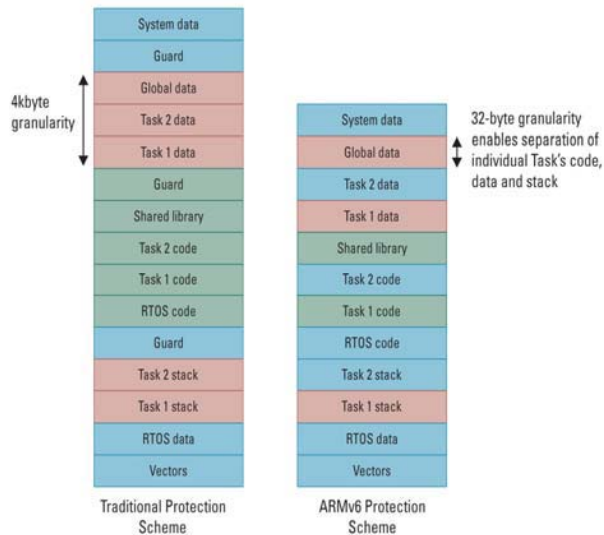

Figure 2. ARMv6 Architecture MPU

**3.1.2 Enhancing Predictability**

Predictability is important in many deeply embedded applications. For example, in automotive, implementing the 'tooth-to-spark' function requires regular and timely action from the processor if it is to be serviced predictably and reliably. The ARM1156T2(F)-S core is intended for higher end applications and will typically be clocked in excess of 200MHz, and will include cache memory. A number of enhancements made improve the predictability of the core, even under these operating conditions.

Adding new instructions to enable exception entry and exit gives a higher-level of predictability to cache-based architectures and enables interrupts to be serviced much sooner following the interrupt request.

The use of cache memory in the system architecture can create issues with system predictability, especially when the required data is absent from the cache memory. Performing a multiple load of say ten 32-bit words from a cache line that is typically eight words long, could require up to three lines of cache to be accessed. In the worst case, if the data for those three cache lines is not available, this could result in three cache line misses. The cache contents would then have to be re-loaded from external memory, causing a considerable delay to the processor before it can proceed to the next instruction or enter into the interrupt service routine.

The availability of a low latency interruptible, re-startable load/store multiple instruction helps to overcome this problem. With the interruptible, re-startable load store instruction, an interrupt can be serviced even if the processor is busy dealing with cache line misses.

This facility makes the use of cache-based architecture much more acceptable for real-time systems, assuming that the cost overhead associated with cache memory is also acceptable. This will still be a factor in safety critical systems where often two or more cores are required to facilitate a redundancy check.

A further enhancement to predictability is derived from the Non-Maskable Interrupt (NMI) added to the Fast Interrupt Request (FIQ). This allows the fast interrupt source to be made non-maskable, which is particularly important when a watchdog is used in the system that needs to be serviced at a particular time.

**3.1.3 Managing Soft Errors**

Soft errors occur as a result of cosmic radiation altering the state of the circuit. Unlike hard errors, soft errors are temporary phenomena and can be corrected. Soft errors pose an increasing risk as geometries shrink, and leading-edge processes at 0.13-micron to 90nm and below, are more vulnerable. Because of the density of transistors, cache memory is especially susceptible to soft errors.

The ARM1156T2(F)-S processor supports fault-tolerant RAM for the cache. The cache instruction line is invalidated if it detects an error, forcing the code to be re-loaded. Any error detected in the TAG RAM generates a cache miss. An error in the cache RAM generates a pre-fetch abort. The core generates precise exceptions for data cache. With data errors there is uncertainty whether the error is within the data or whether it is due to the location address generation.

Data cache errors cause a data abort and the software can then fully recover the data cache contents. The processor also supports fault-tolerant RAM for tightly-coupled memory (TCM). The normal mode of operation for a TCM is that it is kept running to feed the processor – the fault-tolerant support allows a stall to occur. The processor is effectively suspended while the error correction process is invoked. This 'hold and repair' scheme is implemented without the need for an interrupt, i.e. it is supported directly from the core.

**3.2 ARM Cortex-M3 Processor Features**

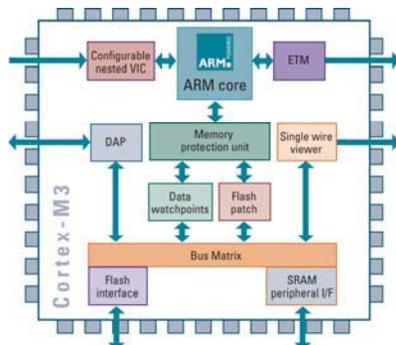

Figure 3: ARM Cortex-M3 Block Diagram

From a performance perspective, many automotive applications do not require the high levels of performance that are achievable from a 32-bit processor. However, many companies would like to standardise on a processor architecture across the automotive platform, to enable reuse of software modules and for the broader economies of scale benefits. Modern vehicles are being developed with an increasing number of processor nodes to facilitate monitoring of sensors for body control, such as electric windows, seat control, oil pressure and temperature monitoring. The capability to network these processors and use them as a multi-core compute machine, distributing and scheduling tasks to processors that have spare capacity, can by realised earlier by standardising on a software architecture. The ARM Cortex-M3 processor, based on Thumb-2 technology, is highly applicable to current 8/16-bit applications that are to be migrated to the ARM architecture.

The ARM Cortex-M3 processor includes specific features to provide highly-efficient support for the less demanding deeply embedded automotive functions.

**3.2.1 Simplified Interrupt Scheme**
Most control systems are heavily loaded with interrupts. A high-end system will typically include many intelligent peripherals that will service some of the low-end interrupt functions, by grouping and buffering – without bothering the processor. With this more intelligent (and expensive) interrupt control process, peripherals can run autonomously for a while. In a lower end system this will not be the case as it adds too much cost to the peripherals. In this case the processor itself will spend a lot of time servicing the interrupts, putting a heavy load on the processing unit. In this situation it is important to have an efficient interrupt response and handling scheme.

In addition, the preamble sequence, which typically saves the system context, and postamble (which restores the system context) to service an interrupt is performed directly in hardware. This means that software developers can write the interrupt service routine in a high-level language and just compile it. On a RISC machine, typically a stub of assembly code would be required at the beginning and end of the interrupt to facilitate the pre and postamble.

Because the pre- and post-ambles are implemented in hardware, parallelism is enabled. It is possible to fetch vectors from the instruction side flash memory while simultaneously writing important system variables. This reduces the number of cycles required to perform an interrupt compared with a scheme managed by software.

The main benefit of this approach is that by aggregating the post and preamble stages, back-to-back handling of interrupts is possible. The processor checks to see if another interrupt is pending at the end of the first interrupt service routine, without restoring the context, in order to get them both serviced in the minimum amount of time.

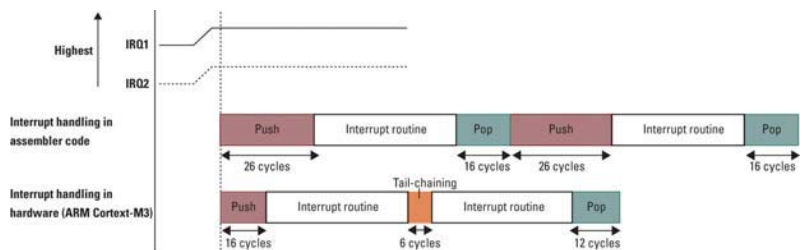

Figure 4: ARM Cortex-M3 Fast Interrupt Response

### 3.2.3 Bit Banding

In all embedded systems, cost dictates that on-chip RAM is a scarce commodity. Consequently, the RAM that is available must be utilised as efficiently as possible.

This means that even simple semaphores (i.e. single bit of data) must be packed into the relevant bits within a memory byte – typically eight semaphores per byte.

Given this storage constraint for semaphores, there is a well-established method for atomic bit manipulation within traditional RISC systems.

To manipulate a particular semaphore within a byte, and to do it atomically (i.e. without disruption from another task which might change the semaphore) it is necessary to disable interrupts to raise the priority of that task, then mask and modify the semaphore and write back to the byte location, before re-allowing interrupts again at the correct priority level.

With the ARM Cortex-M3 processor, up to 1MByte of the memory address region can be mapped, or aliased, to 8MBytes of bit-specific locations elsewhere in the memory map. In this scheme a byte is addressing a particular bit. Because the memory is aliased, it is possible to read and write particular bytes to the RAM normally. If a read/write to the bit addressing alias within the 8Mbyte region is made, it automatically performs a bit-set or bit clear at the aliased address. This enables atomic manipulation by operating on the aliased address. In this scheme, what was a multiple operation task becomes a simple, single write saving many cycles. Now the semaphore can be set or cleared by a single write operation to the aliased byte. No interrupt disabling or bit masking is required.

Bit banding reduces the code required to write the bit to memory, and the core progresses to the next instruction faster. With this scheme any one operation only affects one bit in the byte, so there is now no need to disable other interrupts.

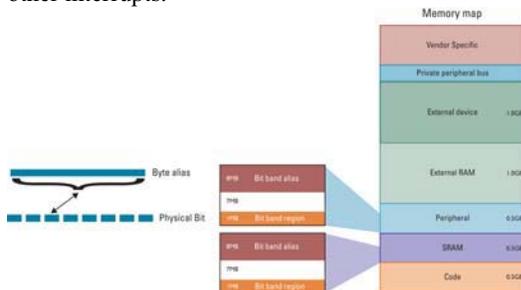

Figure 5: Bit Banding using Aliased Memory

### 3.2.2 Single Wire Debug

Often, low-cost microprocessor solutions are characterized by simple, low pin count packages – typically consisting of just 16 or 32 pins. JTAG, the standard test interface port, is a 5-pin interface. When implemented on a low-pin count device this may represent an unacceptable overhead.

The ARM Cortex-M3 processor includes a single-wire viewer and debug access port - effectively emulating the JTAG standard over a single wire.

Support for an on-the-fly flash memory patch process means that up to eight words can be configured as RAM, providing an equivalent of eight breakpoints in the system. The ability to perform a dynamic download makes access to embedded flash easy, enabling use during the calibration phase, and for writing system and scaling parameters for sensor measurement. The eight discrete breakpoints can be located anywhere, or grouped together as a data sequence.

## 4 Conclusions

To address the technical needs of the automotive industry for embedded software platforms requires a balance of very high performance function with low-cost and extremely efficient implementation. ARM's approach has been to develop two cores on a common instruction set architecture – Thumb-2, which provides the basis for both high performance and code densities.

At the performance end of the spectrum, the ARM1156T2(F)-S processor has been designed to meet the needs of OSEK 2.1.1 RTOS, support fault-tolerant memory interfaces and provide more predictable use of cache memory.

In contrast, the ARM Cortex-M3 processor is intended to support low pin count, low cost devices that are simple to use and easy to design with. By providing efficient support for fast interrupts and atomic bit manipulation, the design of the memory system can be kept clean and very cost-effective.

By enabling the use of a common programming model, the benefits of designing within a unified architecture, such as code reuse and hardware standardisation, become a reality. The tangible benefits to the manufacturer include lower unit cost production, on-time delivery, and improved quality. The vision for the future is to enable the distributed network of processors that is becoming the standard automotive platform to be harnessed as a single compute resource.